\documentclass[12pt,preprint]{aastex}
\newcommand{\vinf}{\mbox{$v_{\infty}$}}

\newcommand{\mdot}{\mbox{$\dot{M}$}}

\shorttitle{Gravity Darkening and Stellar Disks}
\shortauthors{Brown et al.}

\begin{document}

\title{The Effect of Rotational Gravity Darkening on  Magnetically
Torqued Be Star Disks}

\author{J. C. Brown,\thanks{Send offprint requests to:
john@astro.gla.ac.uk(JCB); li@astro.gla.ac.uk(QL) } D. Telfer,
$^1$ Q. Li,$^{1,4}$ R.
Hanuschik,$^2$ J. P. Cassinelli,$^3$ and A. Kholtygin$^{5,6}$\\
$^1$Dept. of Physics and Astronomy, University of Glasgow, Glasgow,
G12 8QQ, UK;\\
$^2$European Southern Observatory, Karl-Schwarzschild-Str. 2, 85748
Garching, Germany; \\
$^3$Dept. of Astronomy, University of Wisconsin-Madison, USA;\\
$^4$Dept. of Astronomy, Beijing Normal University, China;\\
$^5$Astronomical Institute, St.~Petersburg University, Saint
Petersburg State University, V.V.Sobolev Astronomical Institute,
198504 Russia;\\
$^6$Isaac Newton Institute of Chile, St.Petersburg Branch\\
}

\date{Accepted . Received ; in original form}

\begin{abstract}
In the magnetically torqued disk (MTD) model for hot star disks, as
proposed and formulated by \citet{cassi02}, stellar wind mass loss
was taken to be
uniform over the stellar surface.
Here account is taken of the fact that as stellar spin rate $S_o$
($=\sqrt {\Omega_o^2 R^3/GM}$) is increased, and the stellar equator
is gravity darkened, the equatorial mass flux and terminal speed are
reduced, compared to the poles, for a given total $\mdot$.
As a result, the distribution of equatorial disk density, determined
by the impact of north and southbound flows, is shifted further out
from the star. This results, for high $S_o$ ($\gtrsim 0.5$), in a
fall in the disk mass and emission measure, and hence in the observed
emission line EW, scattering polarization, and IR emission.
Consequently, contrary to expectations, critical rotation $S_o \to 1$
is not the optimum for creation of hot star disks which, in terms of EM for
example, is found to occur in a broad peak around  $S_o\approx
0.5-0.6$ depending slightly on the wind velocity law.

The relationship of this analytic quasi-steady parametric MTD model
to other work on magnetically guided winds is discussed.
In particular the failures of the MTD model for Be-star disks alleged
by \citet{owo03} are shown to revolve largely around
open observational tests, rather in the basic MTD physics, and around
their use of insufficiently strong fields.
\end{abstract}
\keywords {stars: emission-line, Be -- stars: magnetic fields --
stars: mass loss -- stars: rotation -- stars: winds, outflows --
polarization.}

\section{Introduction}

Be stars are defined as `non-supergiant B-type stars whose spectra have,
or had at one time, one or more Balmer lines in emission' \citep{coll}. The
pioneering research on Be stars by Struve proposed a rotational
model with emission lines from equatorial disks \citep{struv}, but Be
star disks remain enigmatic despite many decades of
detailed observations and research \citep{jasc,slet82,under,slet87,slet88,smith,port03}. The main 
physics problems they pose are how
the material in them (a)
is delivered from the star; (b) becomes so dense; (c) acquires such
high angular momentum. The answer
to (a) undoubtedly lies in stellar radiation pressure.  The first
qualitative answer to (b) was the Wind Compressed Disk (WCD) model of
\citet{bjor}. In
this, the angular momentum of rotating wind flow returns the matter
to the equator where north and south streams collide and create a
shock compressed disk. There are
several snags with this model. Firstly, it does not produce high
enough densities. Secondly, the disk formed has mainly radial flows
rather than the quasi-Keplerian
azimuthal flows observed. Thirdly, non-radial line-driving forces
\citep{owo96} may cause polar rather than equatorial flow to dominate.

A phenomenological solution to these problems was proposed and
quantified parametrically in the Magnetically Torqued Disk (MTD)
model of \citet{cassi02}.
This invokes a dipole-like field which steers the wind flow toward
the equator and torques up its angular momentum on the way. The field
torques up the
wind flow to Keplerian speeds or higher and confines the radial flow
redirecting it to be poloidal and creating a shock compressed equatorial
disk. The isothermal
disk grows in thickness (but not in density) over comparatively long
timescales ($\approx$ years) which are roughly consistent with long
time-scale variability of some Be stars \citep{doaz,dach,okaz,telt},
allowing a quasi-steady treatment. There is
growing observational evidence of reasonably strong fields (hundreds 
of Gauss) in
hot stars --- e.g.,
$\omega$~Orionis with $B\sim 530 \pm 230$~G \citep{nein},
$\beta$~Cephei with $B\sim  360\pm 30$~G \citep{dona01},
$\theta^1$~Orionis~C with $B\sim 1100\pm100$~G \citep{dona02},
though some of these are very oblique and/or slow rotating and the
MTD model not directly applicable in its basic form.

In the MTD treatment the stellar wind mass flux and wind speed
were taken to be uniform over the stellar surface. For the case of a
rotating star, essential to creating a
disk, this assumption is invalid. Rotation results in equatorial
gravity darkening which reduces the wind mass flux and speed there,
as described by \citet{owo98}. In this
paper, we evaluate the effects of this on the MTD model. We do
so by generalising the basic quasi-steady parametric approach of MTD,
but discussing in Section 5
issues concerning the properties of that description in relation to
other theoretical and observational work on the problem. The MTD
paper was really the first
to model the combined effects of field and rotation but several
earlier papers had discussed disk formation by magnetic channeling
\citep{matt}, while work subsequent to MTD
has variously challenged it \citep{owo03},
and supported it \citep{mahe}.
Insofar as there remains a degree of
disagreement in the literature over whether the model works for Be
stars, our detailed results should be treated with some caution,
though the general trends of the
effect of gravity darkening should be sound.

\section[]{Effect of Gravity Darkening on Disk Density}

\citet{owo98} showed that the mass flux from a stellar surface satisfies
\begin {equation}
F_{mo}(g)\propto g,
\end {equation}
while the terminal speed satisfies
\begin {equation}
\vinf(g) \propto g^{1/2},
\end {equation}
where $g$ is the local effective gravity.

On a rotating sphere with stellar radius $R$ and angular velocity
$\Omega_o$, at colatitude $\theta$, the nett gravity is
\begin {eqnarray}
g(\theta)=\frac {GM}{R^2}- \frac {(\Omega_o R \sin \theta)^2}{R \sin
\theta} \sin \theta        \nonumber     \\
=\frac {GM}{R^2}- \Omega_o^2 R \sin^2 \theta       \nonumber     \\
=g_o (1-\frac {(\Omega_o R)^2}{GM/R} \sin^2 \theta)   \nonumber   \\
=g_o (1-S_o^2 \sin^2 \theta),
\end {eqnarray}
where $S_o=\sqrt {\Omega_o^2 R^3/GM}$ as in MTD. Strictly speaking,
we should also include the effect of continuum scattering radiation
pressure at least, which results in $g_o=\frac
{GM}{R^2}(1-\Gamma_{rad})$, where $\Gamma_{rad}=\frac {L}{L_{Edd}}$
is the ratio of the stellar luminosity to the Eddington luminosity
\citep{maed}.

It follows that the mass flux at $\theta$ becomes
\begin {equation}
F_{mo}(\theta)=K(1-S_o^2\sin^2\theta),      \label {mf1}
\end {equation}
where $K$ is a constant, and the terminal speed for matter from $\theta$ is
\begin {equation}
\vinf(\theta)=\vinf_o(1-S_o^2 \sin^2 \theta)^{1/2},     \label {vinfty}
\end {equation}
where $\vinf_o$ is the value of $\vinf$ at $\theta=0$.

Here we will assume the wind velocity obeys $v_w(r)=\vinf (1-\frac
{R}{r})^{\beta}$, but with
\begin {equation}
v_w(r,\theta)=\vinf(\theta)(1-\frac{R}{r})^{\beta},     \label {vwind}
\end{equation}
where $\beta$ is assumed  not to depend on $\theta$. This basically
requires that wind acceleration occurs quite near the star and that
the field lines are roughly radial there. We want to express $K$ in
terms of the total mass loss rate $\mdot$
\begin {eqnarray}
\mdot=\int ^{\pi}_0 F_{mo}(\theta)\,2\pi R \sin \theta\, R d\theta
\nonumber   \\
=4\pi R^2 K \int^{\pi/2}_0 (1-S_o^2 \sin^2\theta) \sin \theta d\theta
\nonumber   \\
=4\pi R^2 K(1-\frac {2S_o^2}{3}).
\end {eqnarray}

\begin{figure}
\plotone{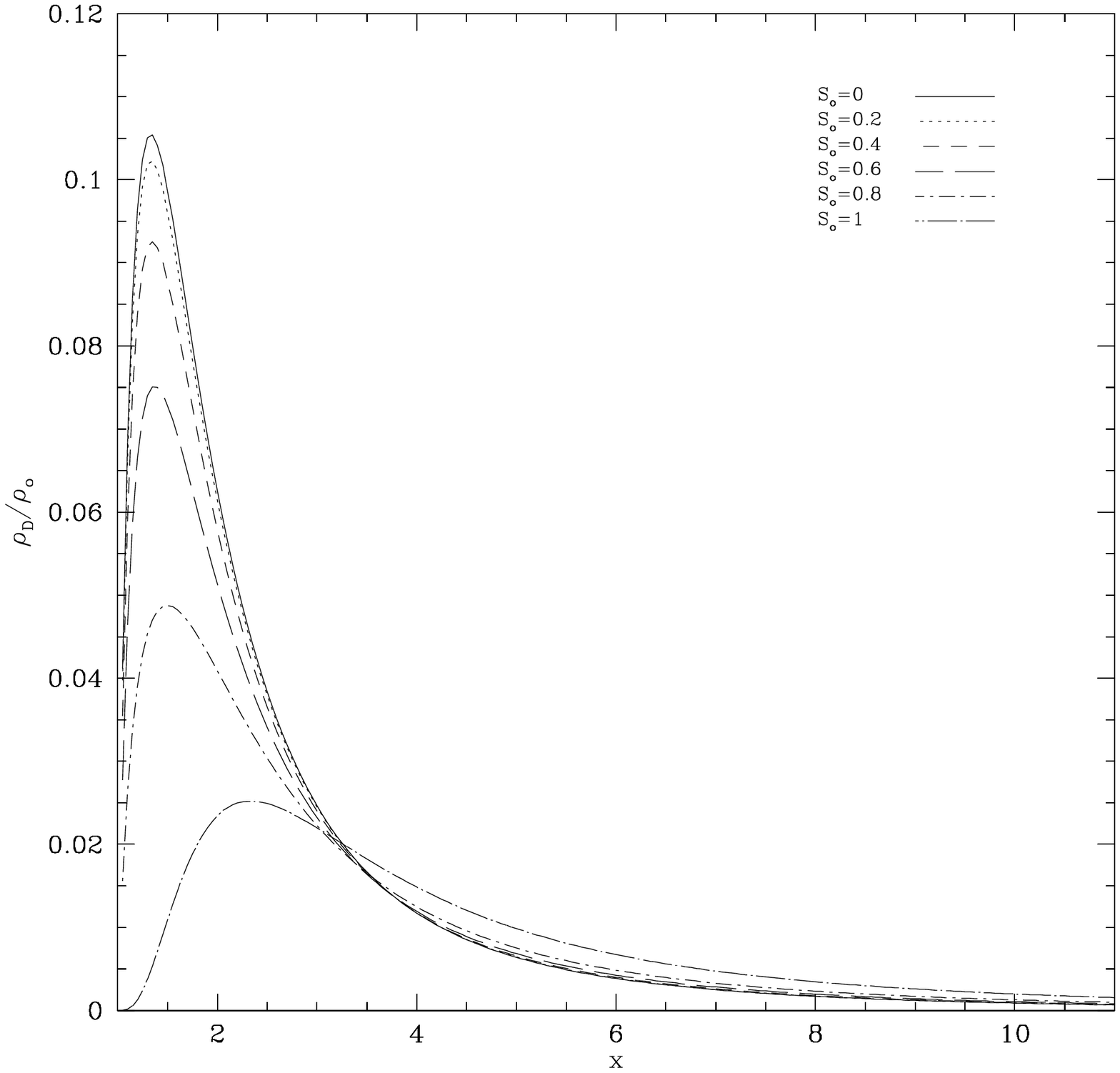}
\caption{Disk density $\frac {\rho_D}{\rho_o}$  vs equatorial
distance $x$ for various spin rates, $S_o$.  Disk density increases
rapidly then decreases sharply again.  The peak value decreases and
moves out as $S_o$ increases, but the peak  is in the range $x\approx
1.3 - 2.3$ for all $S_o$.    \label{rhodf}}
\end{figure}

\begin{figure}
\plotone{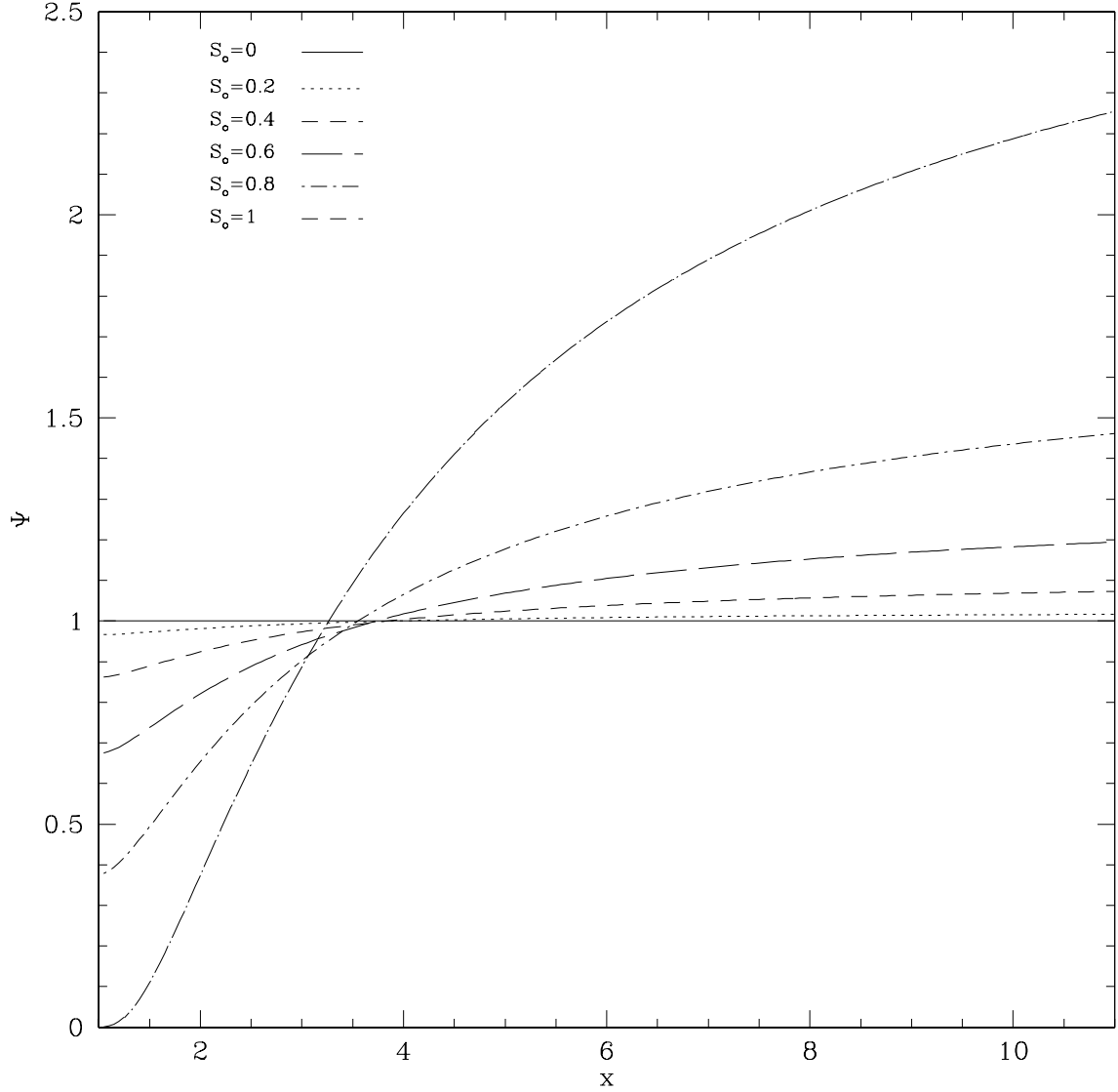}
\caption{Ratio of disk density to that without gravity darkening
$\Psi =\frac {\rho_D(S_o,x)}{\rho_D(0,x)}$ vs equatorial distance $x$
for various spin rates $S_o$.       \label{psif}}
\end{figure}

So by equation (\ref{mf1}),
\begin {equation}
F_{mo}(\theta)=\frac {\mdot}{4\pi
R^2(1-2S_o^2/3)}(1-S_o^2\sin^2\theta).   \label {mf2}
\end {equation}
To relate this mass flux at $\theta$ on the stellar surface to that
normal to the equatorial plane at distance $x=r/R$, we follow
\citet{cassi02} in parametrizing the decline of magnetic field, $B$, with
equatorial plane distance according to
\begin {equation}
B(x)=B_o x^{-b},                        \label {magnetic}
\end {equation}
where $B_o$ is taken as uniform over, and normal to, the stellar
surface, and $b$ is a constant with a dipole field $b=3$. Flux
conservation then requires that a cross-sectional area $dA$ of a flux
tube arriving at the equator is related to its area $dA_o$ at the
star by $B dA=B_o dA_o$ or
\begin {equation}
dA(x)=dA_o\,x^b.       \label {area}
\end {equation}
However, $dA(x)=2\pi R^2 xdx$ and $dA_o=2\pi R^2 \sin \theta d\theta$
so that the relation between $\theta$ and $x$ is given by integrating
$\sin \theta d\theta=x^{-b+1} dx$, to yield
\begin {equation}
\cos \theta =C-\frac {x^{-b+2}}{b-2}.
\end {equation}
Requiring that $x \to \infty$ for $\theta \to 0$ implies $C=1$ so that
\begin {equation}
\cos \theta(x)=1-\frac {x^{-b+2}}{b-2},
\end {equation}
For the particular case $b=3$ (the dipole case) mainly discussed by
\citet{cassi02}, and which we focus on henceforth, this yields
\begin {equation}
\cos \theta (x)=1-\frac {1}{x}.      \label {cos}
\end {equation}
We can now obtain the mass flux $F_m(x)$ at $x$ near the equatorial
plane by using $F_m(x)=F_{mo}(\theta)dA_o(\theta)/dA(x)$ and using
equations (\ref {mf2}), (\ref {area}) and (\ref {cos}), namely
\begin {equation}
F_m(x)=\frac {\mdot}{4\pi R^2 (1-2S_o^2/3)} x^{-3}
\{1-S_o^2[1-(1-\frac {1}{x})^2]\},
\end {equation}
while by equations (\ref {vinfty}), (\ref {vwind}) and (\ref {cos}),
the wind speed there is
\begin {equation}
v_w (x)=\vinf_o(1-\frac {1}{x})^{\beta}\{1-S_o^2[1-(1-\frac {1}{x})^2]\}^{1/2},
\end {equation}
and the arriving wind ram pressure $P_{ram}(x)=F_m(x)v_w$ is
\begin {equation}
P_{ram}=\frac {\mdot \vinf_o}{4\pi R^2}x^{-3}(1-\frac {1}{x})^{\beta}
\frac{\{1-S_o^2[1-(1-\frac {1}{x})^2]\}^{3/2}}{1-2S_o^2/3}.
\label {ramp}
\end {equation}
The high density $\rho_D(x)$ of cool, shock - compressed, disk (WCD)
in the equatorial plane is then given as in the WCD model (and in
MTD) by the isothermal disk (sound speed $c_s$) pressure balance
expression $\rho_D c_s^2=P_{ram}=P_D$ or by equation (\ref{ramp})
\begin {equation}
\rho_D (S_o,x)=\rho_o x^{-3} (1-\frac
{1}{x})^{\beta}\frac{\{1-S_o^2[1-(1-\frac
{1}{x})^2]\}^{3/2}}{1-2S_o^2/3}.              \label {rhod}
\end {equation}
Here
\begin {equation}
\rho_o=\frac {\mdot}{4\pi R^2 \vinf_o}(\frac {\vinf_o}{c_s})^2
\label{rhoo}
\end {equation}
which is equivalent to $\rho_{Dc}$ in the MTD model.
Based on equation (\ref {rhod}), we can write the disk density
allowing for rotational gravity darkening, compared to that
neglecting it (e.g., MTD) as
\begin {equation}
\Psi(S_o,x)=\frac{\rho_D(S_o,x)}{\rho_D(0,x)}=\frac {[1-\frac
{S_o^2}{x}(2-\frac {1}{x})]^{3/2}}{1-2S_o^2/3}       \label {psix}
\end {equation}
with the property $\Psi(S_o,1)=(1-S_o^2)^{3/2}/(1-2S_o^2/3)$ which is
$<1$ for all $S_o$, and tends to 0 as $S_o \to 1$. This is because
the equatorial wind flow falls with increasing $S_o$, reducing the
inner disk compression. On the other hand, if we (formally --- see
comments below in Section 3) apply equation (\ref {psix}) as $x \to
\infty$ we would get for the disk behavior
\begin {equation}
\Psi(S_o,\infty) \rightarrow \frac {1}{1-2S_o^2/3}.       \label {psiinfty}
\end {equation}
This is always $>1$ because the polar wind supply of mass to large,
equatorial distances $x$, is increased (for fixed $\mdot$) for large
$S_o$. Also, as $S_o \rightarrow 1$ (critical rotation) we find
\begin {equation}
\Psi (1,x)=3(1-\frac {1}{x})^3
\end {equation}
which is $>1$ at $x>3^{1/3}/(3^{1/3}-1)\approx 3.26$.
In Figs.~\ref{rhodf} and \ref{psif}, we show $\frac {\rho_D}{\rho_o}$
and $\Psi(S_o,x)$ versus $x$ for various $S_o$.
Fig.~\ref{rhodf} shows the disk density to peak in the range
$x\approx 1.3 - 2.3$ for all $S_o$, then to decrease rapidly with $x$
for all $S_o$.

\section {Effect of Gravity Darkening on Disk Extent, Mass, and
Emission Measure}
We have seen that  $\rho_D(x)$ decreases  and moves its maximum
somewhat to larger $x$ values as $S_o$ increases. However, we need
also to consider the extent of the disk, i.e., the lower and upper
boundaries of equation (\ref{rhod}) in $x$ as limited by the magnetic
field strength. While enhancement of $\rho_D$ locally enhances the
local contribution per unit volume ($\sim \rho_D^2$) to the disk
emission measure, it makes the material there harder to torque so
that the extent of the disk is modified. In particular, for example,
equation (\ref {psiinfty}) is not valid in practice since the rapid
decline in $B(x)$, as $x$ goes up, limits the torquing to a finite
distance.

To estimate the effect of including rotational gravity darkening on
observable disk properties, we need to assess the effect on the inner
and outer disk boundaries. Here we do so using a somewhat simpler
treatment than that in MTD, namely what MTD termed the `switch
approximation'. In this, the disk is taken to be rigidly torqued by
the magnetic field (i.e., $v=v_o x$, where $v_o=S_o \sqrt {GM/R}$)
out to the distance where the magnetic energy density $\frac
{B^2}{8\pi}$ falls below the rotational kinetic energy density
$U_{KE}=\frac {1}{2}\rho_D v^2$. We have then, by equation (\ref
{magnetic}),
\begin {equation}
U_B=\frac {B^2}{8\pi}=\frac {B_o^2}{8\pi}x^{-6}
\end {equation}
and by equation (\ref {rhod})
\begin {equation}
U_{KE}=\frac {1}{2} \rho_o \frac {GM}{R}S_o^2 x^{-1} (1-\frac
{1}{x})^{\beta} \frac {\{1-S_o^2[1-(1-\frac
{1}{x})^2]\}^{3/2}}{1-2S_o^2/3}.
\end {equation}
So the outer disk boundary $x=x_{outer}(S_o,\gamma)$ is given by
setting $U_B=U_{KE}$. Thus $x_{outer}$ is the solution $x$ to
\begin {equation}
x^5(1-\frac{1}{x})^{\beta}\frac{\{1-S_o^2[1-(1-\frac
{1}{x})^2]\}^{3/2}}{1-2S_o^2/3}=\frac {\gamma^2}{S_o^2},      \label
{xouter}
\end {equation}
where
\begin {equation}
\gamma=(\frac {B_o^2/8\pi}{GM\rho_o/2R})^{1/2}         \label{gamma}
\end {equation}
is a measure of field energy compared to the disk gravitational energy.
The inner disk boundary in the present approximation is simply the
Keplerian rotation distance (c.f., MTD)
\begin {equation}
x_{inner}=S_o^{-2/3}.             \label {xinner}
\end {equation}

\section {Results and Discussion}
\begin{figure}
\plotone{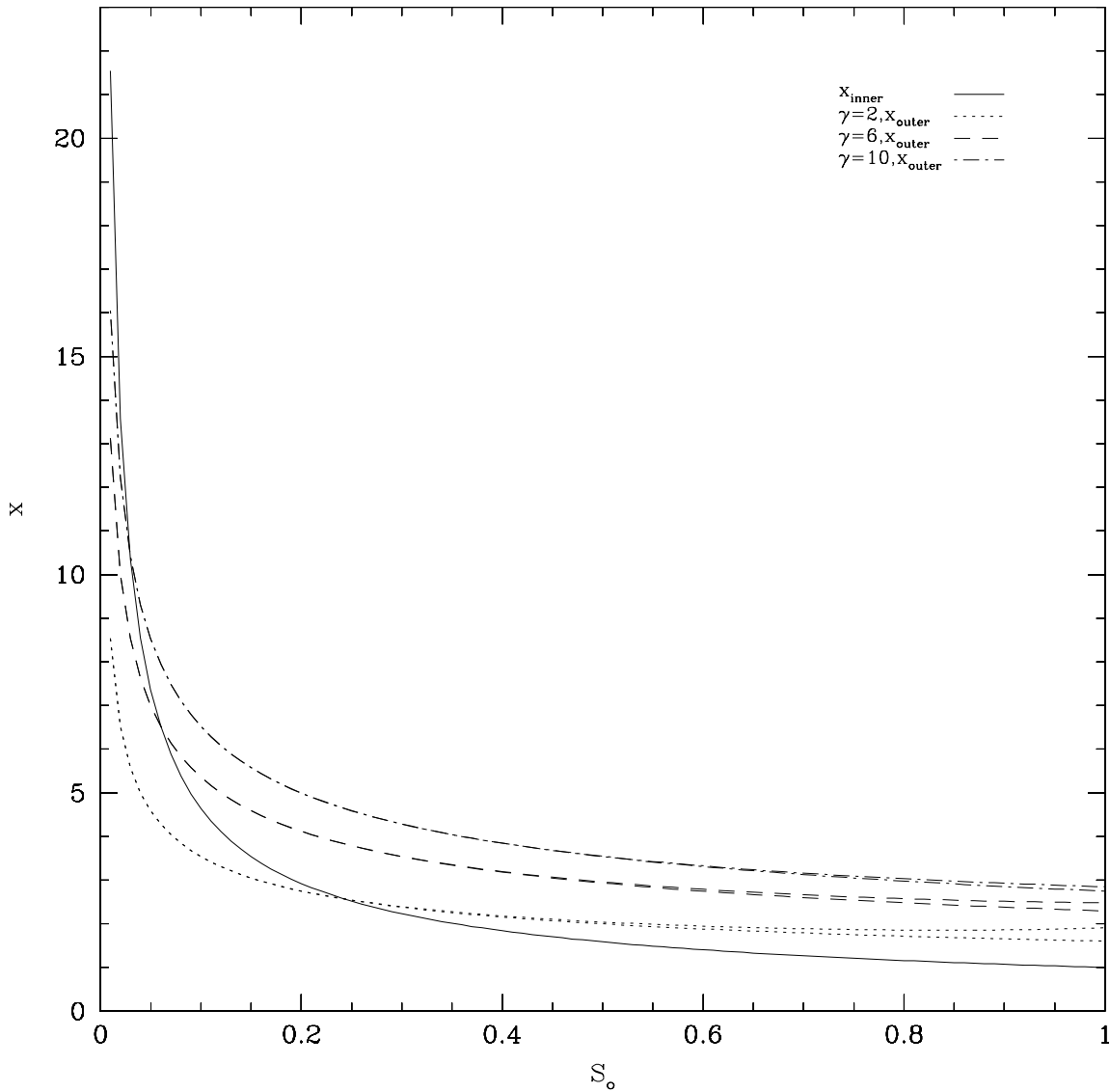}
\caption{Disk boundaries $x_{inner}$ and $x_{outer}$ vs $S_o$ for
various  $\gamma$. The solid line  corresponds to the inner radius
$x_{inner}$, and the other lines to the outer radius of the disk for
various $\gamma$. Each pair of curves with the same  line type gives
results  with gravity darkening (upper curve) and without gravity
darkening (lower curve).  For larger $\gamma$, one requires smaller
$S_o$ in order for a disk to form ($x_{outer} > x_{inner}$). This
implies that for small magnetic fields, a relatively fast spin rate
$S_o$ is necessary, while for large magnetic fields, moderate $S_o$
is adequate.  Comparing curves with and without gravity darkening
shows that gravity darkening has only a minor effect on the outer
radius of the disk for any given $\gamma$, $S_o$.      \label{xxf}}
\end{figure}

\begin{figure}
\plotone{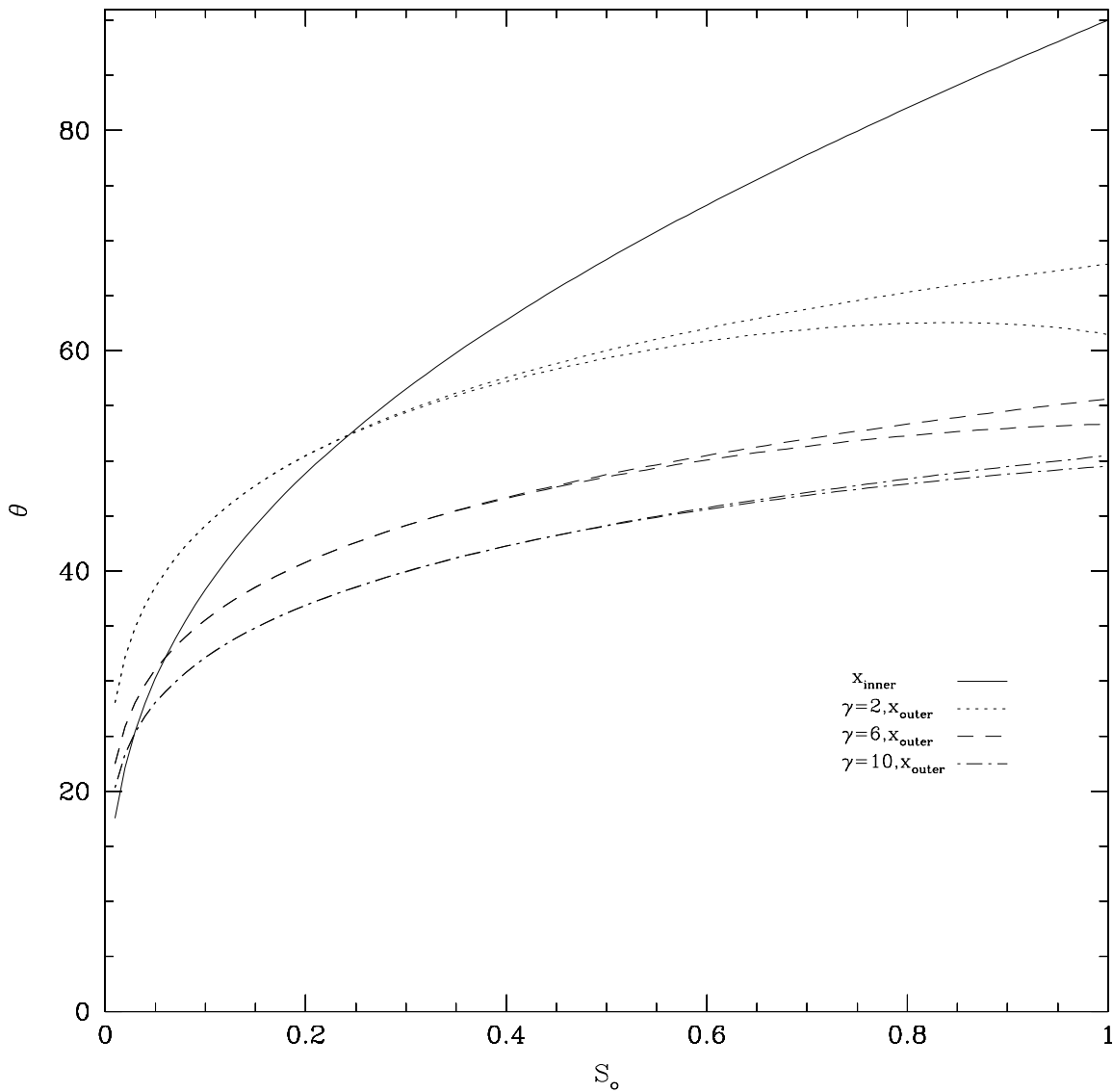}
\caption{Stellar surface colatitudes $\theta^o$ corresponding to the
inner and outer disk boundaries, vs $S_o$ for various  $\gamma$ . The
solid line  corresponds to the inner boundary, and the other lines to
outer boundaries of the disk. Each pair of curves with the same  line
type  gives results with gravity darkening (lower curve) and without
gravity darkening (upper curve).  The possible range of colatitudes
channeling matter are angles between the curve of solid line type and
the other curve for a given $\gamma$.       \label{thetaf}}
\end{figure}

In Figs.~\ref{xxf} and \ref{thetaf}, we show $x_{inner}(S_o)$ and
$x_{outer}(S_o)$ versus $S_o$ for various $\gamma$ values,
and the corresponding colatitudes, on the stellar surface,
of $x_{inner}$ and $x_{outer}$  in terms of equation (\ref{cos}).
It turns out
that these boundaries do not change greatly with $S_o$ once $S_o$ is larger
than 0.2 --- 0.3, but change a lot with
$\gamma$, and that the mass flux reaching the disk comes from a rather small
range of colatitudes (e.g. for $S_o=0.6$, $45^o\lesssim \theta \lesssim 70^o$
with $\gamma=6$ --- see Fig.~\ref {thetaf}). Mass flow from the pole
(small $\theta$) leaves the  star as part of the wind while equatorial flow
(large $\theta$) does not achieve Keplarian speed.

According to equation (\ref{gamma}), $\gamma$ is determined by the
magnetic and gravitational fields. In order to make mass flux
channeling and torquing possible, $\gamma$ has to be substantially
greater than unity.   In terms of  observations, the magnetic fields
of Be stars are no larger than  hundreds of Gauss. Hence, $\gamma$
should probably be in the range of $1 \to 10$  for Be stars to meet
this requirement.  In Figs.~\ref{xxf} and \ref{thetaf}, we also see
that for smaller $\gamma$, a larger $S_o$ is necessary for a disk
(clearly, the outer radius of the disk must be larger than the inner
radius). These two figures also show that the gravity darkening has a
small effect on the outer radius of the disk, which is within $\sim 5
R$ for appropriate $\gamma$ and $S_o$. Similar outer boundaries have
been derived for some stars using different disk models by \citet{doug}
and \citet{cote}.

The detection of disks by polarization, infrared emission, and
emission line strength is related to their mass and their emission
measure which are proportional to $\int_V\rho_D dV$ and
$\int_V\rho_D^2 dV$, respectively, where $V$ is the disk volume.
If the disk has thickness $H(x)=h(x)R$ at distance $x$ then it
contains a total number of particles
\begin {equation}
N=\frac {2\pi R^3}{m} \int_{x_{inner}(S_o)}^{x_{outer}(S_o,\gamma)}
\rho_D(x)h(x)xdx,
\end {equation}
and has emission measure
\begin {equation}
EM=\frac {2\pi R^3}{m^2}
\int_{x_{inner}(S_o)}^{x_{outer}(S_o,\gamma)} \rho_D^2(x)h(x)xdx,
\end {equation}
where $m$ is the mean mass per particle. Using equation (\ref{rhod})
these yield
\begin {equation}
N=N_o \int_{x_{inner}(S_o)}^{x_{outer}(S_o,\gamma)} \frac
{x^{-2}h(x)}{1-2S_o^2/3}(1-\frac
{1}{x})^{\beta}\{1-S_o^2[1-(1-\frac{1}{x})^2]\}^{3/2}dx,     \label
{number}
\end {equation}
where $N_o=2\pi R^3\frac {\rho_o}{m}$,
and
\begin {equation}
EM = EM_o \int_{x_{inner}(S_o)}^{x_{outer}(S_o,\gamma)} \frac
{x^{-5}h(x)}{(1-2S_o^2 /3)^2}(1-\frac {1}{x})^{2\beta}
\{1-S_o^2[1-(1-\frac{1}{x})^2]\}^3dx,    \label {emissionmeasure}
\end {equation}
where $EM_o=2\pi R^3 (\frac {\rho_o}{m})^2$.

\begin{figure}
\plotone{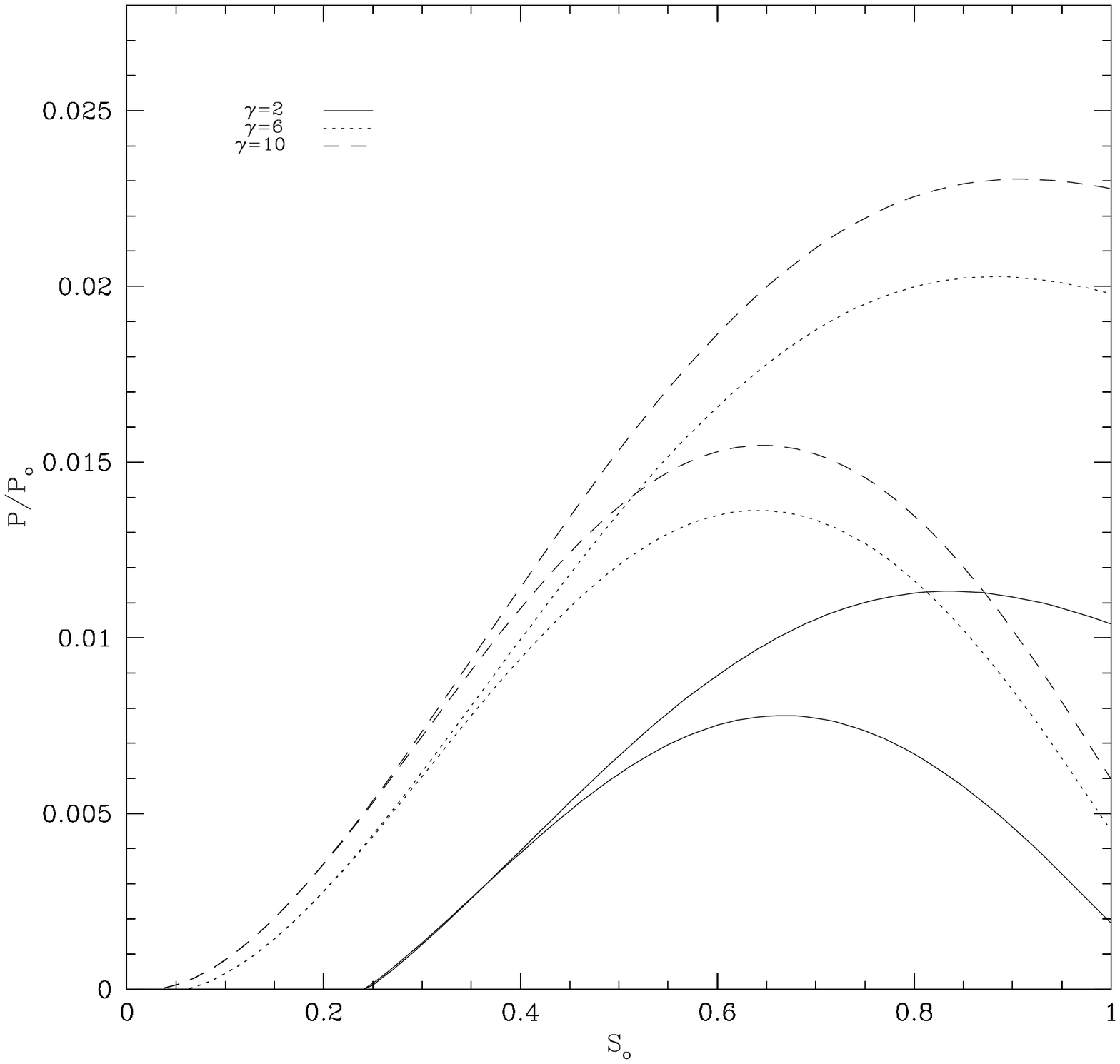}
\caption{Ratio of polarization $\frac {P}{P_o}$ relative to arbitrary
reference value $P_o$,  vs $S_o$ for various  $\gamma$.  The
inclination angle is assumed to be $90^o$ (edge-on observation).
Lower and upper curves with the same line type are for gravity
darkening and no gravity darkening, respectively.
\label{polf}}
\end{figure}

\begin{figure}
\plotone{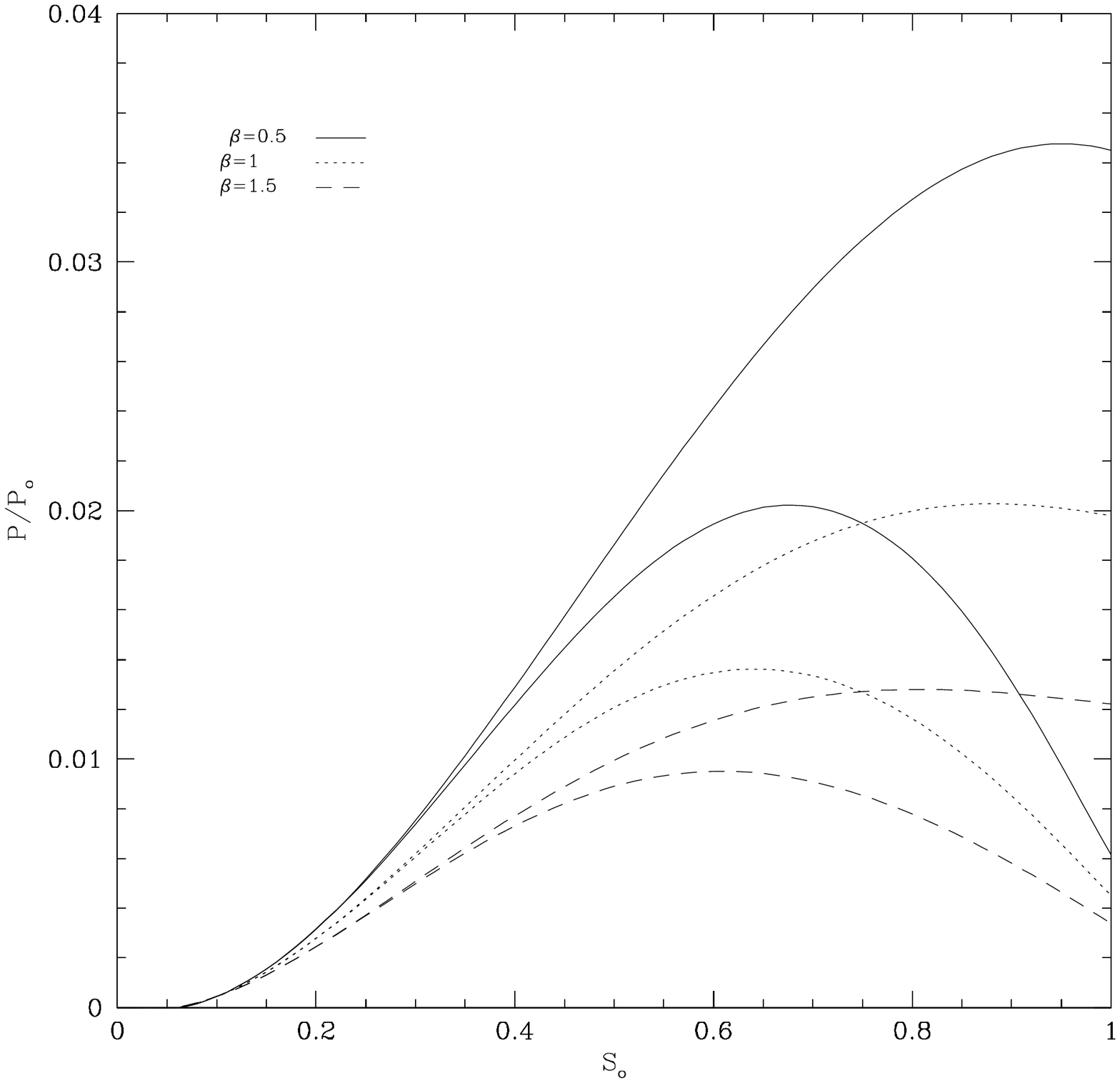}
\caption{Ratio of polarization $\frac {P}{P_o}$ relative to arbitrary
reference value $P_o$, vs $S_o$  for various  $\beta$ with
$\gamma=6$.  The inclination angle is assumed to be $90^o$ (edge-on
observation).  Lower and upper curves of the same line type are for
gravity darkening and no gravity darkening, respectively.
\label{polbetaf}}
\end{figure}

Following the \citet{brown77} formulation, the scattering polarization is
$P=\tau (1-3\Gamma)\sin^2 i$, where $\tau$ is optical depth, $\Gamma$
is the shape factor of the
disk and $i$ is the inclination angle. Assuming the disk to be a slab
with constant thickness $H=Rh$ and including the finite source
depolarization factor $D=\sqrt
{1-R^2/r^2}=\sqrt{1-1/x^2}$  \citep{cassi87,brown89}, then we have the optical
depth $\tau$,
   \begin {eqnarray}
\tau=\frac {3\sigma_T}{16}
\int^{r_2}_{r_1}\int^{\mu_2}_{\mu_1}n(r,\mu)D(r)dr\,d\mu \nonumber
\\
=\frac {3\sigma_T R}{16} \int^{x_{outer}}_{x_{inner}}\int^h_0 n(x,z)
D(x) \frac{x}{x^2+z^2} dx\,dz   \nonumber \\
=\tau_o
\int^{x_{outer}}_{x_{inner}}x^{-3}(1-\frac{1}{x})^{\beta}\frac{[1-S_o^2
(\frac{2}{x}-\frac{1}{x^2})]^{3/2}}{1-2S_o^2/3}\sqrt{1-\frac{1}{x^2}}\arctan\frac{h}{x}dx, 
\label {tau}
\end {eqnarray}
where $\tau_o=\frac {3\sigma_T R}{16}\frac{\rho_o}{ m}$, $\sigma_T$
is Thomson cross section, $n=\rho_D/m$ is the electron density of the
disk, and $\mu$ is the cosine
of the angles between the incident light to the disk and the
rotational axis. As in MTD, we neglect the absorption and suppose a
fully ionized disk.  The shape factor
$\Gamma$ yields
\begin {eqnarray}
\Gamma=\frac {\int^{r_2}_{r_1}\int^{\mu_2}_{\mu_1}n(r,\mu)D(r) \mu^2
dr\,d\mu}{\int^{r_2}_{r_1}\int^{\mu_2}_{\mu_1}n(r,\mu)D(r)dr\,d\mu}
\nonumber \\
=\frac {\int^{x_{outer}}_{x_{inner}}n(x)D(x) [\frac{1}{2x}\arctan
\frac{h}{x}-\frac{h}{2(h^2+x^2)}]xdx}{\int^{x_{outer}}_{x_{inner}}n(x)D(x)\arctan\frac{h}{x}dx}. 
\label {Gamma}
\end {eqnarray}
Substituting equations (\ref{tau}) and (\ref{Gamma}) in the original
polarization expression and after some reduction, yields the
polarization, $P$, with gravity darkening effects,
\begin{equation}
P=P_o I_P                        \label{pol}
\end {equation}
where $P_o=\tau_o=\frac{3\sigma_T R \rho_o}{16m}$ and $I_P$ is the integral
\begin{equation}
I_P=\int^{x_{outer}}_{x_{inner}}x^{-3}(1-\frac{1}{x})^{\beta}\frac{[1-S_o^2
(\frac{2}{x}-\frac{1}{x^2})]^{3/2}}{1-2S_o^2/3}\sqrt{1-\frac{1}{x^2}}
\arctan\frac{h}{x}dx (1-3\Gamma )\sin^2 i.
\end{equation}
Then $I_P=P/P_o$ is found numerically and shown in Figs.~\ref{polf}
and \ref{polbetaf} for $h=0.5$ with respect to typical half opening
angles of the disk about $10^o$, (e.g., \citet{hanu};
\citet{port96}), and the inclination angle $i=90^o$.

\begin{figure}
\plotone{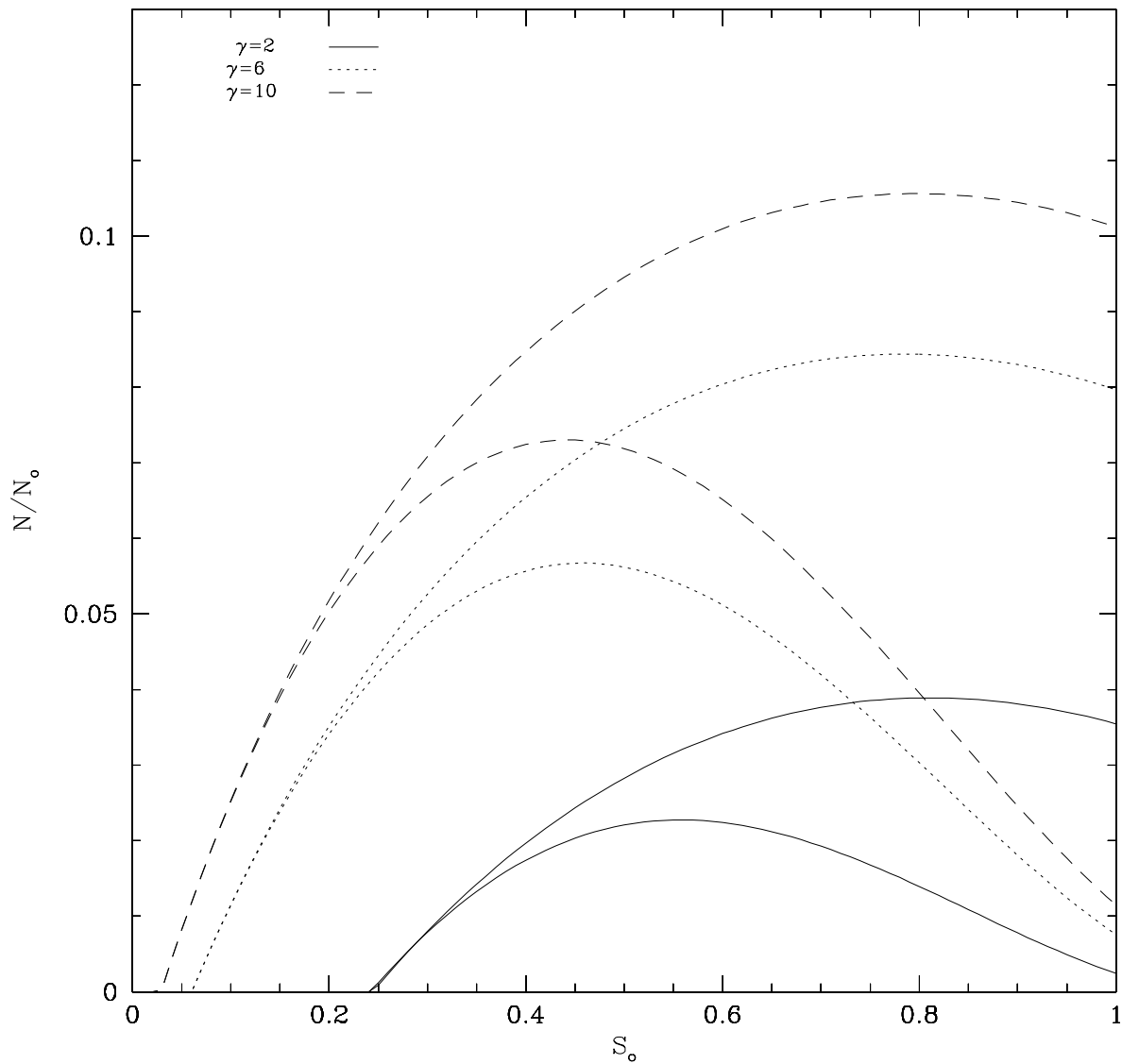}
\caption{Ratio of total number of particles in the disk $\frac
{N}{N_o}$ relative to arbitrary reference value $N_o$, vs $S_o$ for
various  $\gamma$.  Lower and upper curves with the same line type
are for gravity darkening and no gravity darkening, respectively.
Gravity darkening significantly decreases the results as $S_o$
becomes large, and produces a peak. For larger $\gamma$, a smaller
$S_o$ gives rise to the formation of a disk, so the curve becomes
wider. The larger $\gamma$, the bigger $\frac {N}{N_o}$.
\label{partif}}
\end{figure}

\begin{figure}
\plotone{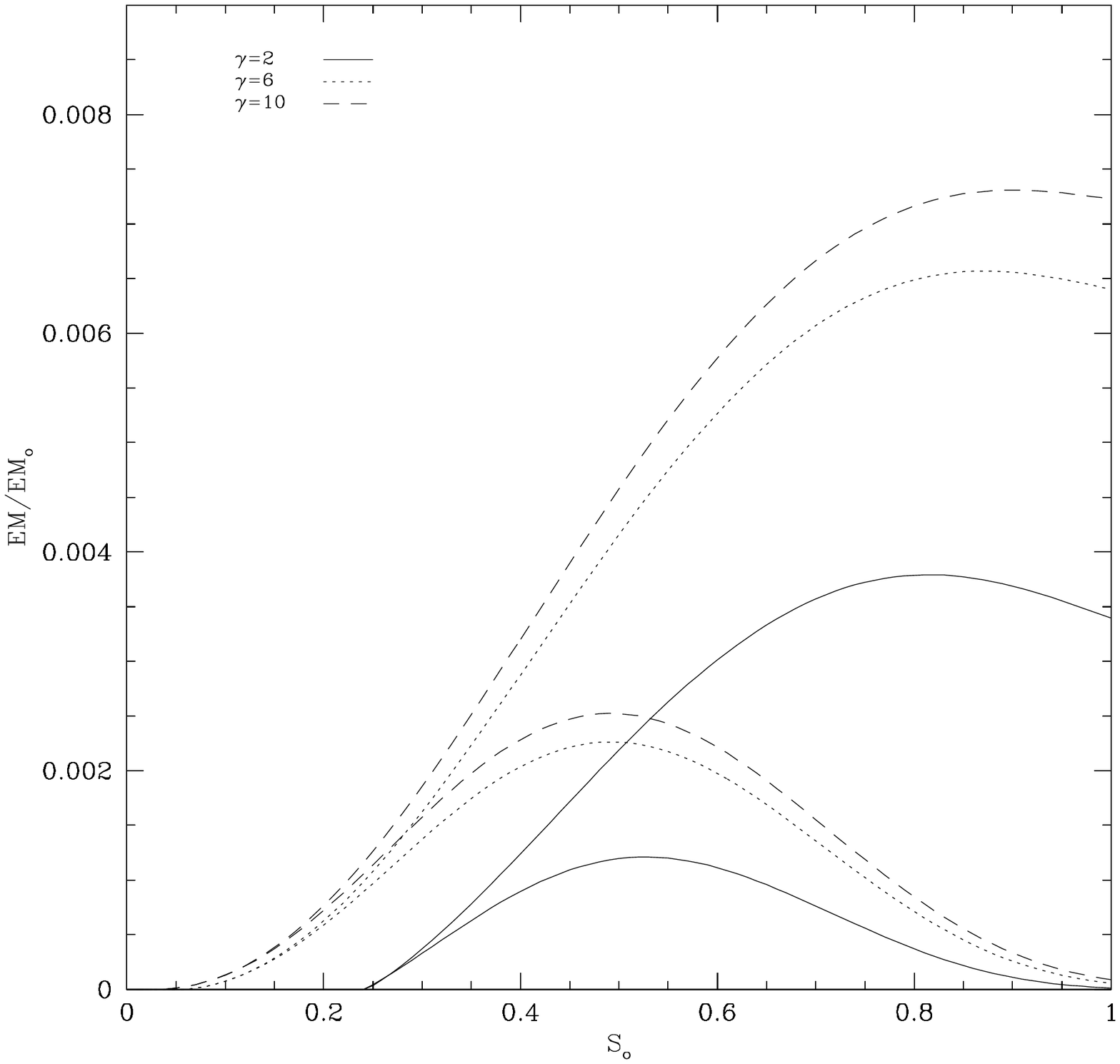}
\caption{Ratio of emission measure $\frac {EM}{EM_o}$ relative to
arbitrary reference value $EM_o$, vs $S_o$ for various  $\gamma$.
Lower and upper curves with the same line type are for gravity
darkening and no gravity darkening, respectively.       \label{emf}}
\end{figure}

In Figs.~\ref{partif} and \ref{emf}, we show the results of equations
(\ref{number}) and (\ref{emissionmeasure}) for $N(S_o,\gamma)$ and
$EM(S_o,\gamma)$ together with those obtained when gravity darkening
is ignored (using same switch approximation). For the latter, we use
the integrands as in equations (\ref{number}) and
(\ref{emissionmeasure}), but with $S_o=0$; the same lower limit
$x_{inner}=S_o^{-2/3}$ as given by equation (\ref{xinner}); and the
outer limit the solution to equation (\ref{xouter}), with $S_o=0$ on
the left. We see that increasing $S_o$ from zero results in a rising
disk mass and emission measure up to a broad maximum at $S_o\lesssim
0.5$ and falling back almost to 0 as $S_o \to 1$.  It is not
surprising to see that gravity darkening has strong effects on total
number of particles and emission measure, since gravity darkening
effects significantly reduce the mass flow from equatorial stellar
regions into the disk. We also note that $N/N_o$ and $EM/EM_o$ ratios
have peaks at about $S_o\approx 0.5$ for all $\gamma$, while for no
gravity darkening they essentially keep increasing. Similar results
are shown in Fig.~\ref{polf} for the polarization which depends
mainly on the disk electron scattering mass. The more the total
number of particles  in the disk, the stronger the polarization.  If
gravity darkening were neglected,  one would get a large $I_P$ so a
small $P_o$, for a given observed polarization value $P$ (equation
\ref{pol}).  This would imply a smaller $\rho_o$ since $P_o\propto
\rho_o$, which implies an underestimation of the mass loss rate
$\mdot$ (equation \ref{rhoo}), if we ignore gravity darkening.

\begin{figure}
\plotone{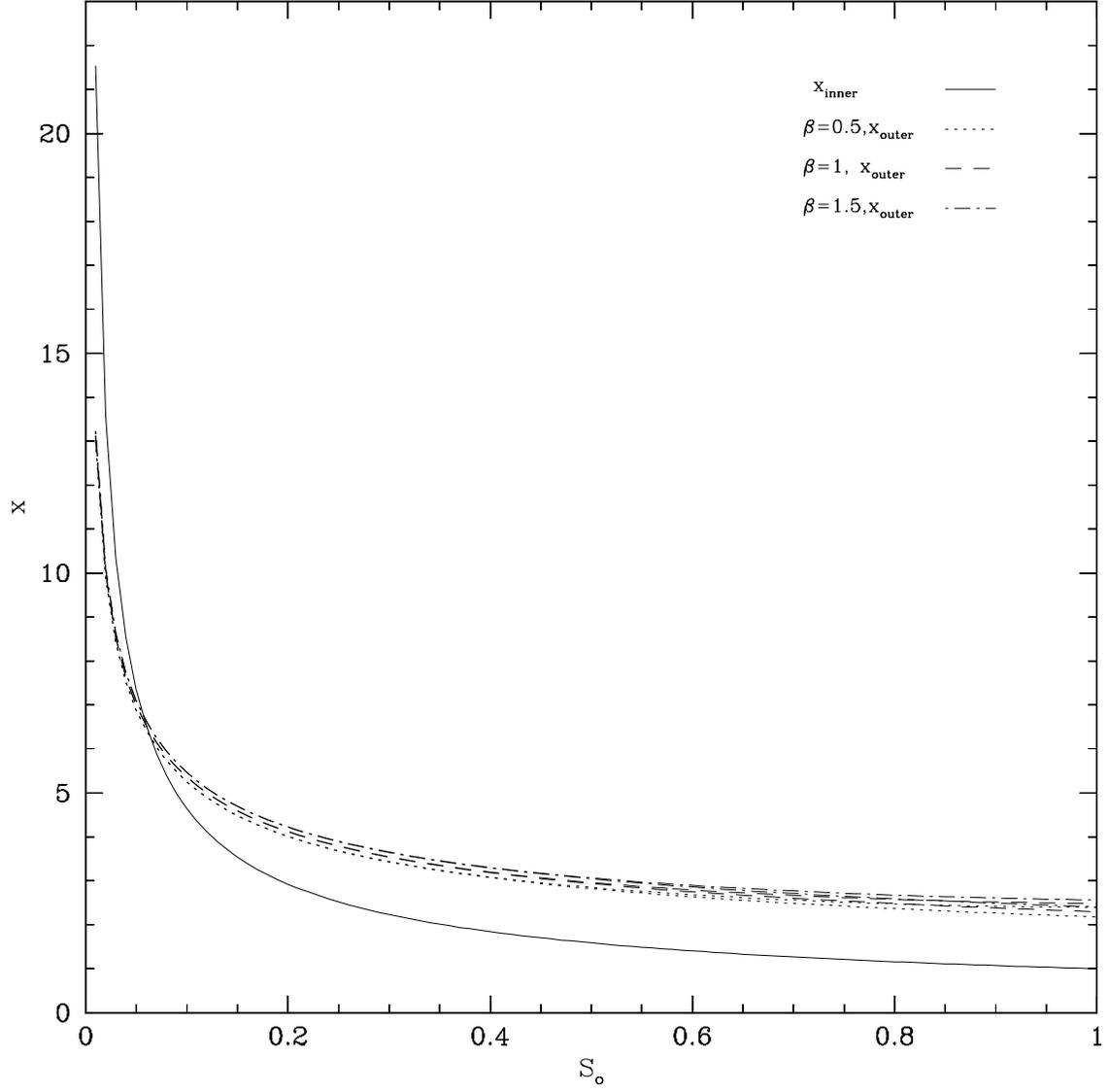}
\caption{Disk boundaries $x_{inner}$ and $x_{outer}$ vs $S_o$ for
various  $\beta$ with $\gamma=6$. The solid line  corresponds to the
inner boundary $x_{inner}$, and the other lines to outer boundaries
of the disk.  Lower and upper curves of the same line type are for
gravity darkening and no gravity darkening, respectively.
\label{xxbetaf}}
\end{figure}

\begin{figure}
\plotone{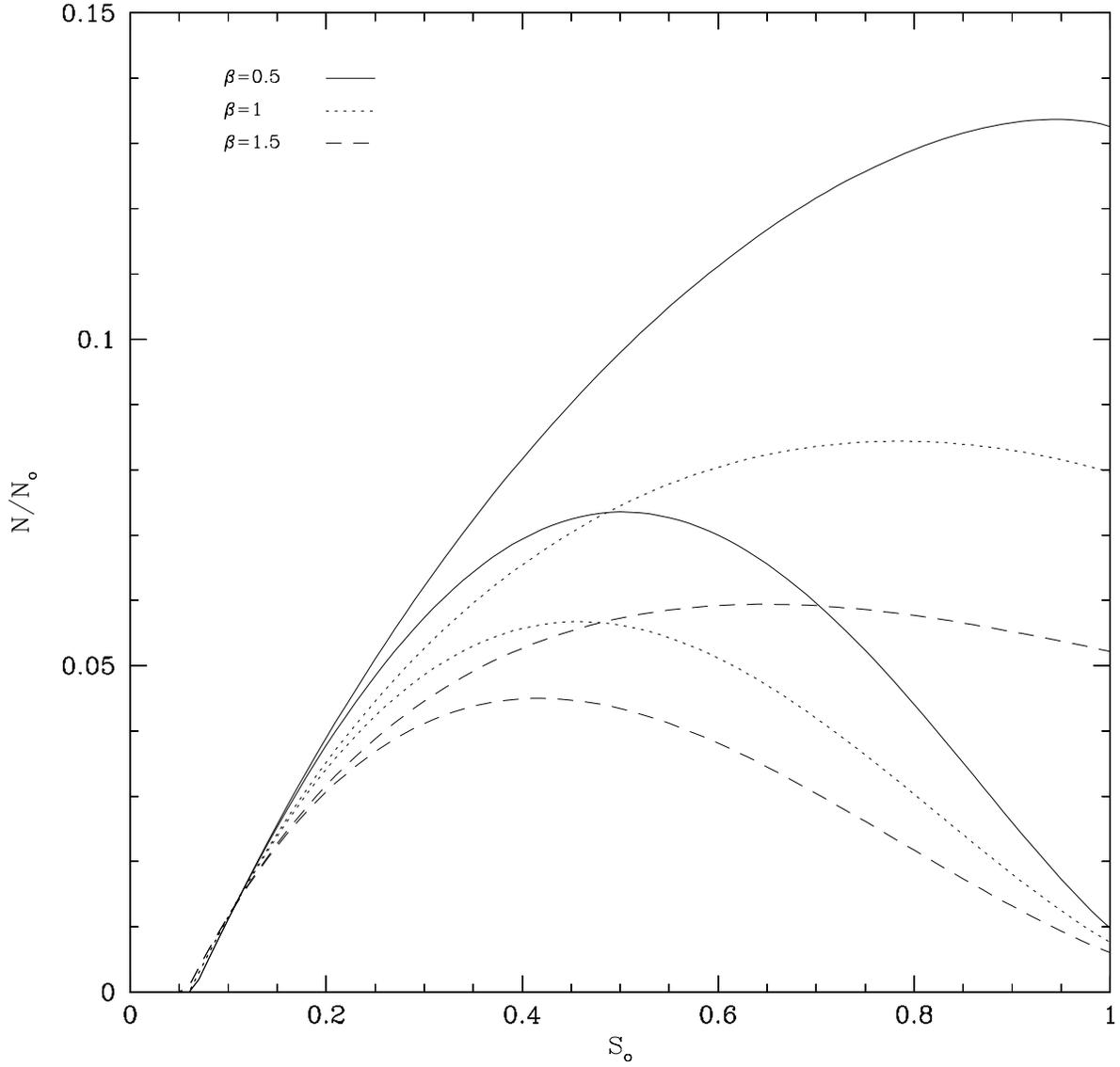}
\caption{Ratio of total number of particles $\frac {N}{N_o}$ relative
to arbitrary reference value $N_o$, vs $S_o$  for various  $\beta$
with $\gamma=6$.   Lower and upper curves of the same line type are
for gravity darkening and no gravity darkening, respectively.
\label{partibetaf}}
\end{figure}

\begin{figure}
\plotone{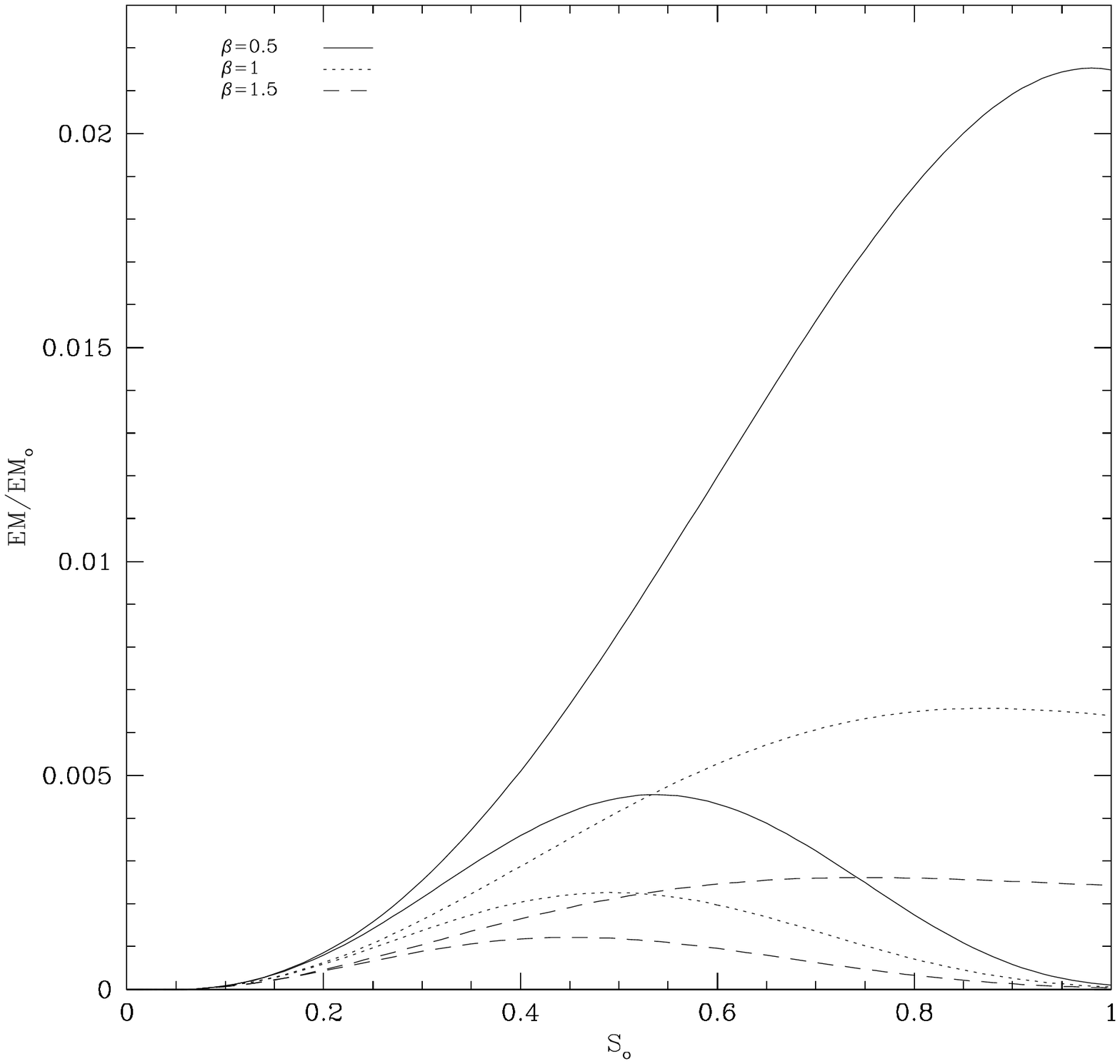}
\caption{Ratio of emission measure $\frac {EM}{EM_o}$ relative to
arbitrary reference value $EM_o$, vs $S_o$ for various  $\beta$ with
$\gamma=6$.   Lower and upper curves of the same line type are for
gravity darkening and no gravity darkening, respectively.
\label{embetaf}}
\end{figure}

The previous treatment is for fixed $\beta=1$. In order to see the
influence of the velocity law on the results, we tried various
$\beta$ values for a given $\gamma$. Fig.~\ref {xxbetaf} shows that
wind velocity law has minor effects on the disk boundaries and so
does gravity darkening. Figs.~\ref{partibetaf}, \ref{embetaf} and
\ref{polbetaf} show that slower winds (i.e., bigger
$\beta$ values) will lead to much smaller total number of particles,
emission measure and polarization of the disk, as well as gravity
darkening significantly decreases these disk properties. From the
plots we see that, for small $\beta=0.5$, the total
number of particles, emission measure and polarization peaks shift
slightly to larger rotation rate ($S_o\approx 0.6$), while a
statistical study of observation data
indicates the most common $S_o\approx 0.7$ \citep{port96}
which, in our interpretation, would favor small $\beta$,
i.e., fairly fast acceleration of winds from the
stellar surface. If individual disk detectability peaks
for $S_o \approx 0.6$ and actual
detection peak for $S_o \approx 0.7$, either there is a
bias/selection effect in operation \citep{town},
or there in an upward trend in
the frequency distribution of $S_o$ values.

\section{Comparison of MTD with MHD Simulations and Other Work on
Magnetic Channeling}
It was noted in Section 1 that the MTD model is phenomenological
and parametric, and not a full solution to the physics equations. It
is aimed, like all such models,
at describing the main features of a system accurately enough to
reproduce the essential physics but simple enough to facilitate ready
incorporation of additional effects
(such as gravity darkening) and comparisons with data. It is of
course important to evaluate how well the MTD model describes the
reality when compared with more complete
solutions. A full and detailed comparison is beyond the scope of this
paper but we summarise here the present status as we see it of the
relation of MTD to recent analytic
and numerical work on closely related problems.

One of the earliest studies of the problem which found disk formation 
was that in a `magnetospheric'
context was by \citet{havn}. \citet{kepp99,kepp00} carried out
numerical
simulations of magnetised stellar winds with rather weak fields
and found disk `stagnation zones' in
the equatorial plane. \citet{matt}
studied non-rotating winds in dipole fields and found persistent
equatorial disk structures around AGB stars though with a steady
throughput of mass
leaking through the disk.  \citet{mahe}
conducted a detailed
analytic study of the MTD situation and
found that persistent disks are formed for quite small fields though
he obtained somewhat tighter constraints than MTD on the relevant
regimes of magnetic field and spin
rate. In the MHD simulations of isothermal flow
driven outward from a non-rotating star with dipole magnetic
fields, \citet{doul02}
found that the effect of magnetic fields in channeling stellar winds
depends on the overall ratio of magnetic to flow kinetic energy density,
(as did MTD) and obtained disk results with rather low fields.

In contrast to all of these, in the \citet{doul03}
conference paper, based on the same code as \citet{doul02},
the interpretation shifts somewhat and seems more negative about disk
persistence. \citet{owo03} added rotation to their earlier work and
concluded that no stable disk
could form, with matter either falling back or bursting out after a
modest number of flow times. In fact, the MHD code they used is 
incapable of handling
the larger fields which MTD argued were required and which are recently in
fact observed (several hundred Gauss) in some Be stars, so their 
numerical results are not that relevant. Furthermore, given that 
observed fields are strong enough so that the wind is bound to be 
steered and torqued to the equatorial regions, if the behaviour of 
that matter were highly unstable as Ud-Doula and Owocki's simulation results, we should 
observe very frequent Be star disk disruption, which we do not. In fact we 
see no reason why material should fall back, given that it is 
centrifugally supported, until the dipole structure fills up. This 
takes a very large number of flow times (many years) - hence the the 
quasi-steady formulation in MTD.
In the case of weak fields and low rotation, fall back and burst out 
of matter is not altogether
surprising, but it is not clear why the \citet{owo03}
simulation results conflict with those of others. For stronger fields
using order of magnitude scaling estimates, they found that disks,
essentially like MTD can form and have gone on to develop scenarios
for strong fields closely akin to MTD under
the names Magnetically Rigid Disk and Magnetically Confined Wind
Shocked Disk. They were, however, dismissive of the relevance of
this to Be-stars. This was
not on the grounds of the physics of MTD but over the issue of
whether a semi-rigid disk near co-rotation
can be reconciled with observations of Be-stars, specifically
spectrum line shapes and the long term V/R variations. The work of
\citet{telf} suggests that the
former is not a serious problem. The issue of the V/R variations was
emphasised in the original MTD paper  which recognised that, if MTD
is the correct description of Be
disk formation, a close examination is required of how the V/R
variations could arise. At first sight it would seem that the
conventional interpretation in terms of
spiral density waves (induced by the non--spherical potential) in a
Keplerian disk would not work if the field controlled the disk and
another interpretation would have to
be found but it depends on the rate of viscous
diffusive redistribution of disk matter toward Keplerian
\citep{mahe}. Until further
testing is carried out we are therefore of the view that the MTD model remains
a good basic scheme for further modelling work. We also note that,
when MTD is applied across the
range of hot star spectral types, it offers a remarkably good
explanation of the narrow spectral range where disks are in fact
detected. No other model offers any explanation of this observation.

\section{Conclusions}
We have discussed the phenomenological magnetically torqued disk
(MTD) model of \citet{cassi02} for hot
(particularly Be) star disk formation in
relation to other work on magnetically steered wind creation of disk
like structures, concluding that, for moderate fields comparable to
those observed,
the description is physically realistic but that further work is
needed to see if its disk velocity structure can be reconciled with
observations including line
profiles and V/R variations. We have recognised that the basic model
did not recognise the effect of spin induced gravity darkening on the
latitudinal distribution of
wind flow and consequently on disk density structure. We have
included this effect and found that, although increasing $S_o$  from
zero favors disk formation, at high
$S_o$ the polar shift in mass flux results in decreasing disk
detectability  by emission or polarization. The fact that
detectability (say above half of the height of the
peak values of emission and polarization in Figs.~\ref{emf}, \ref{polf}, 
\ref{embetaf}, and \ref{polbetaf}) covers quite broad rotation
range in $S_o$, namely  $S_o
\approx 0.25 - 0.80$, is generally in good agreement with the fact that Be
star rotation rates are typically estimated to occur most frequently
near 0.7 (e.g., \citet{port96}, and references therein; \citet{yudin}),
which tends to favour fast
acceleration velocity laws. Overall this means that in the MTD model,
the most easily
observable disks are, contrary to naive expectation, not expected
from the fastest rotators, but from moderate ones, as observed,
though we note the
\citet{town} comments on the effect of gravity darkening in inferring
underestimated line width rotation rates. Clearly the whole MTD
scenario needs further work
to test it thoroughly, including reconciliation of the phenomenology,
numerical MHD, and analytic MHD theoretical treatments and work on
further diagnostics
such as X-ray emission from the MTD deceleration shocks of Be stars
for comparison with $ROSAT$ and other X-ray datasets.

\section*{Acknowledgements}
The authors acknowledge support for this work from: U.K. PPARC
Research Grant (JCB); NASA grant TM3-4001A (JPC,JCB); and Royal Society
Sino-British Fellowship Trust Award (QL), NSFC grant 10273002 (QL); 
RFBR grant 01-02-16858
(AK). An anonymous referee's comments lead to a
significant improvement of the paper.


\end {document}